\documentclass{bioinfo}
\copyrightyear{2013}
\pubyear{2013}

\usepackage{natbib} 
\usepackage[breaklinks]{hyperref} 
\usepackage{url}
\usepackage{breakurl}
\usepackage{graphicx}
\usepackage{xspace}
\usepackage{algorithm}
\usepackage{algorithmic}
\usepackage{comment}
\usepackage{subfigure}
\usepackage{color}

\newcommand{\kmers}{$k$-mers\xspace}
\newcommand{\kmer}{$k$-mer\xspace}
\newcommand{\ourtool}{{\sc KmerGenie}\xspace}

\newcommand{\aureus}{\emph{S. aureus}\xspace}
\newcommand{\hchr}{\emph{chr14}\xspace}
\newcommand{\bombus}{\emph{B. impatiens}\xspace}

\newenvironment{revision}{\par}{\par}
\newcommand{\revisedtext}[1]{{#1}}

\title{Informed and Automated  {\em k}-Mer Size Selection for  Genome Assembly}
\author[Chikhi and Medvedev]{Rayan Chikhi$^{1}$ and Paul Medvedev$^{1,2}$}
\address{$^{1}$Department of Computer Science and Engineering, The Pennsylvania State University, USA.\\
$^{2}$Department of Biochemistry and Molecular Biology, The Pennsylvania State University, USA.
}
\date{Mar. 2013}
\history{Received on XXXXX; revised on XXXXX; accepted on XXXXX}
\editor{Associate Editor: XXXXXXX}

\begin{document}

\maketitle

\begin{abstract}
\textbf{Motivation:} 
Genome assembly tools  based on the de Bruijn graph framework rely on a parameter $k$, 
which represents a trade-off between several competing effects that are difficult to quantify.
There is currently a lack of tools that would automatically estimate the best $k$ to use and/or
quickly generate histograms of \kmer abundances that would allow the user to make an informed decision.

\textbf{Results:} 
We develop a fast and accurate sampling method that constructs approximate abundance histograms with
a several orders of magnitude performance improvement over traditional methods.
We then present a fast  heuristic that uses the generated abundance histograms for putative $k$ values to estimate the best possible value of $k$.
We test the effectiveness of our tool using diverse  sequencing datasets and find that its choice of $k$ leads to some of the best assemblies.

\textbf{Availability:} Our tool \ourtool is freely available at: \\ \url{http://kmergenie.bx.psu.edu/}

\textbf{Contact:} chikhi@psu.edu
\end{abstract}
\section{Introduction}

Genome assembly continues to be a fundamental aspect of high-throughput sequencing data analysis.
In the years since the first methods were developed, there have been numerous improvements and the field is now rich with tools that provide biologists 
several options~\citep{Ribeiro13,SOAPdenovo2,spades,minia,IDBAUD,SGA,Velvet}. 
Many of these tools are based on the de Bruijn graph framework, where reads are chopped up into \kmers 
(substrings of length $k$)~\citep{Pevzner01}.
The de Bruijn graph is constructed with  nodes being the ($k-1$)-mers and the edges being the $k$-mers present in the reads.
Broadly speaking, an assembler constructs the graph, performs  various graph simplification steps, and outputs non-branching paths as contigs -- 
contiguous regions which the assembler predicts are in the genome.

Recently, there have been several meta-analyses of assemblers that have pointed to systematic shortcomings of current methods~\citep{Assemblathon1,GAGE,Alkan11,Assemblathon2}.
The Assemblathon competitions~\citep{Assemblathon1,Assemblathon2} demonstrated that assembling a dataset still requires significant expert intervention. 
One issue is many assembler's lack of robustness with respect to the parameters and the lack of any systematic approach to choosing the parameters.
In de Bruijn based assemblers, the most significant parameter is $k$, which determines the size of the \kmers into which reads are chopped up.
Repeats longer than $k$ nucleotides can tangle  the graph and break up contigs, thus, a large value of $k$ is desired.
On the other hand, the longer the $k$ the higher the chances that a \kmer  will have an error in it; 
therefore, making $k$ too large decreases the number of correct $k$-mers present in the data.
Another effect is that when two reads overlap by less than $k$ characters, they do not share a vertex in the graph
and thus create a coverage gap that breaks up a contig. 
Therefore, the choice of $k$ represents a trade-off between several effects.

Because some of these trade-offs have been difficult to mathematically quantify, 
there has not been an explicit formula for choosing $k$ taking into account all these effects.
It is possible to calculate some bounds based on estimated genome size and coverage (e.g. by applying Lander-Waterman statistics), however,
such estimates do not usually take into account the
 impact of repetitiveness of the genome, heterozygousity rate, or read error rate.
In practice, $k$ is often chosen based on prior experience with similar datasets.
More thorough approaches compare assemblies obtained from different $k$ values; however, 
they are very time consuming since a single assembly can take days for mammalian-size genomes.
A more informed initial choice for $k$ can be made by building abundance histograms for putative values of $k$ and comparing them.
The abundance histogram shows the distribution of \kmer abundances (the number of occurrences in the data) for a single $k$ value. 
Such histograms can provide an expert with valuable information for choosing $k$ --
however, the time to construct such a histogram can take up to a day for just a single value of $k$~\citep{DSK,Jellyfish}  

Our contribution in this paper is two-fold.
First, we propose an accurate sampling method that constructs approximate abundance histograms with
an order of magnitude speed improvement, compared to traditional tools based on $k$-mer counting.
Our method allows an expert user to make an informed decision by quickly generating abundance histograms for many $k$ values and analyzing the results, either visually or statistically.
Our second contribution is a fast heuristic method for selecting the best possible value of $k$, based on the generated abundance histograms for many values of $k$.
The heuristic is based on the intuition that the best choice of $k$ is the one which provides the most distinct non-erroneous 
(genomic) \kmers to the assembler.
Our method can be integrated into assembly pipelines so that the choice of $k$ is made automatically without user intervention.

We implement our methods in a publicly available tool called \ourtool.
We test \ourtool's effectiveness using three sequencing datasets from a diverse set of genomes: \aureus, human chromosome 14, and \bombus.
First, we find that our approximation of the histogram is very close to the exact histogram and easily separable from histograms for nearby values of $k$.
Next, we judge the accuracy of \ourtool's choice of $k$ by assembling the data for numerous $k$ values and comparing the quality of the assemblies.
We find that \ourtool's choice leads to 
the best assemblies of \aureus and \bombus, as measured by the contig length (NG50),
and to  a good assembly of \hchr that represents a compromise between contig length and the number of errors.


\section{Methods}
Our method can be summarized as follows.
We start by generating the abundance histograms for numerous putative values of $k$.
We then fit  a generative model to each histogram in order to estimate how many distinct \kmers 
in the histogram are genomic (i.e. error-free).
Finally, we pick the value of $k$ which maximizes the number of genomic \kmers.
We now describe each step in detail.

\subsection{Building the abundance histograms \label{sec:histograms}}

Consider a multiset of reads $R$ from a sequencing experiment. 
For a given value of $k$, each read is seen as a multiset of $k$-mers. 
For instance with $k=3$, the read ATAGATA is the multiset of five $3$-mers (ATA, TAG, AGA, GAT, ATA). 
By taking the union of all reads, a dataset of reads is also seen as a multiset of $k$-mers.
Each \kmer is said to have an abundance, which is the number of times it appears in the multiset.
A common function used to understand the role of $k$ is the abundance histogram.
For a given abundance value $i$, the function tells the number of distinct \kmers with that abundance.


One way to calculate the abundance histogram is to first run a \kmer counting algorithm.
A \kmer counting algorithm takes a set of reads and outputs every present \kmer along with its abundance.
The abundance histogram can then be calculated in a straightforward way.
$K$-mer counting is itself a well-studied problem with efficient solutions and tools, though even efficient implementations can take hours or days on large datasets~\citep{Jellyfish, DSK}.
Such a solution would be inefficient for generating histograms for multiple values of $k$
since one would need to run the \kmer counter multiple times.

Instead, we propose to create an approximate histogram by sampling from the \kmers, 
an idea explored in a more general setting by~\citet{cormode}.
The pseudocode of our algorithm is shown in Algorithm~\ref{alg:histogram}.
Intuitively, we use a parameter $\epsilon$ to dictate the proportion of distinct $k$-mers we sample.
We pick a hash function $\rho_\epsilon : \{A,C,G,T\}^k \rightarrow [0..\epsilon]$ that uniformly distributes the universe of all possible \kmers into $\epsilon$ buckets.
In our implementation, we adopted a state-less 64 bits hash function (RanHash, page 352 in~\citep{NumericalRecipes}).
We then count the abundances of only those  \kmers that hash to 0. 
The abundance histogram is then computed from the \kmer counts, scaling the number of \kmers with a given abundance by $\epsilon$.

\begin{algorithm}
\begin{algorithmic}[1]
\REQUIRE An integer $k>0$, a set of reads $R$, $\epsilon>0$.
\ENSURE Approximate abundance histogram of \kmers in $R$
    \STATE Init empty hash table $T$, with default values of 0.
    \FORALL{\kmers $x$ in $R$}
        \IF{$\rho_\epsilon(x) = 0$}
            \STATE $T[x] = T[x] + 1$
        \ENDIF
    \ENDFOR
    \STATE Compute abundance histogram $h_\epsilon$ of $T$
    \STATE Let $h(i) = \epsilon \cdot h_\epsilon(i)$ for each $i$
    \STATE Output $h$
    
\end{algorithmic}
\caption{Compute approximate abundance histograms}
\label{alg:histogram}
\end{algorithm}

The running time of the algorithm is $O(|R|(\ell - k))$, where $\ell$ is the read length.
The expected memory usage is $O(m / \epsilon)$, where $m$ is the number of distinct \kmers in $R$.
Though the asymptotic running time is the same as for an exact \kmer counting algorithm,
the total overhead of adding \kmers to a hash table is reduced by a factor of $\epsilon$.
Similarly, though the memory usage is asymptotically the same as for an exact \kmer counter,
the decrease by a factor of $\epsilon$ can make it feasible to store the hash table in RAM.

\subsection{Generative model for the abundance histogram }
Given an abundance histogram, our next step is to infer the number of distinct genomic \kmers in it. 
In principle, if we knew the error rate,  we could easily estimate the number of genomic \kmers (not necessarily distinct) as a proportion of the total number of \kmers.
However, such a simple approach does not allow us to estimate which of the \kmers are genomic and hence
does not allow us to estimate the number of {\em distinct} genomic \kmers. 
Moreover, the error rate is itself a parameter that is not known prior to assembly, 
so it must be estimated as well.

Instead, our approach is to take a generative model and fit it to the histogram. 
We can then infer the number of distinct genomic \kmers from the parameters of the model. 
Fitting a model to a histogram has been previously explored in the context of error-correction~\citep{Chaisson,Quake}.
We adopt the model proposed by \citet{Quake}, which we describe here for completeness.

\subsubsection{Haploid model}

The $k$-mer abundance histogram is a mixture of two distributions: one representing  genomic $k$-mers, and one representing erroneous $k$-mers. We use the term \emph{copy-number} to denote the number of times a genomic $k$-mer is repeated in the genome.
A genomic \kmer distribution is itself a mixture of $n$ Gaussians, each Gaussian corresponding to $k$-mers with a copy-number $1\leq i  \leq n$. We fix the maximum copy-number to $n=30$. For each copy-number $i$, the mean $\mu_i$ and variance $\sigma_i^2$ of the Gaussian \revisedtext{are different values (due to Illumina biases)}, proportional to the copy-number. Thus in our model, the mean and variance $(\mu_1,\sigma_1^2)$ (for copy-number $1$) are free parameters, and the remaining means and variances are fixed to $\mu_i=i\mu_1$ and $\sigma_i^2=i\sigma_1^2$. The weights of the mixture of Gaussians are given by a Zeta distribution, which has a single free shape parameter $s$. 
The erroneous \kmers distribution is modeled as a Pareto distribution with fixed scale of $1$ and a free shape parameter $\alpha$. The mixture between erroneous and genomic \kmers is weighted by a free parameter $p_e$, which corresponds to the probability that a \kmer is erroneous. Thus in total, our model has five free parameters ($\mu_1,\sigma_1^2,s,\alpha,p_e$).


\subsubsection{Extension to the diploid model}


In the diploid case, we say that a \kmer is homozygous if it appears in both alleles, and is heterozygous otherwise. 
We model the genomic \kmers of a diploid organism as a mixture of two haploid genomic \kmer distributions $D_{ht}$ and $D_{hm}$. 
A mixture parameter ($p_h$) controls the proportion of homozygous \kmers (drawn from $D_{hm}$) to heterozygous \kmers
(drawn from $D_{ht}$).
The parameters of the two distributions remain free, with the following exception.
As homozygous \kmers are expected to be sequenced with twice the coverage of heterozygous \kmers, the mean of $D_{hm}$ is fixed to twice the mean of $D_{ht}$. 
As in the haploid case, we model the erroneous \kmers as a Pareto distribution with parameter $\alpha$ and 
a mixture proportion of $p_e$.
In summary, the diploid model has eight free parameters: the variances ($\sigma_1^2$) and the Zeta  shapes  ($s$) 
for the $D_{hm}$ and $D_{ht}$ distributions; additionally, the mean ($\mu_1$) of $D_{ht}$, the  Pareto shape ($\alpha$), 
the error probability ($p_e$), and the heterozygosity proportion ($p_h$).



\subsection{Fitting the model to the histogram}

Given the abundance histogram, we can  estimate the parameters of the above model that would be the best
fit for the observed histogram.
Similarly to what is done in \citet{Quake}, we do a maximum-likelihood estimation of the parameters  using the \verb!optim! function in R (BFGS algorithm). 
Let $d$ be the number of distinct \kmers present in the histogram. 
Let $\widehat{p_e}$ be the estimated mixture parameter between the erroneous and genomic \kmers in the above model. 
Immediately, $\widehat{p_e}d$ is an estimate of the numbers of erroneous \kmers and $(1-\widehat{p_e})d$ is an estimate of the number of genomic \kmers present in the reads. 

\subsection{Finding the optimal $k$}

\begin{revision}

Our key insight is that the best value of $k$ for assembly is the one which provides the most distinct genomic \kmers.
To see this, consider the number of distinct $k$-mers in the reference genome. 
Observe that, as $k$ increases, this number also increases and approaches the length of the genome,
 as a consequence of repeat instances becoming fully spanned by $k$-mers. 
Thus, a high number of distinct genomic $k$-mers allows the assembler to resolve more 
repetitions.
In the ideal scenario of perfect coverage and error-free reads, the best value of $k$ for assembly would be the read length. 
However, the read coverage is typically imperfect and reads are error-prone, requiring a more nuanced approach.

First, consider obvious high and low thresholds: for $k$ close to the read length, it is unlikely that all the $k$-mers in the reference are present in the reads due to imperfect coverage. 
On the other hand, note that any genome will contain all $k$-mers for a small enough $k$ 
(e.g. the human genome contains all possible $4$-mers). 
This is due to the fact that a significant chunk of a genome behaves like a random string.


Next, we examine the values of $k$ between these two thresholds. Essentially, two effects are competing. The shorter a $k$-mer is, the more likely it is (i) to appear in the reads, but also (ii) to be repeated in the reference. For a typical sequencing depth and values of $k$ near the read length, it is likely that only a small fraction of the $k$-mers from the reference genome appears in the reads (effect (i)). However, unless the sequencing depth is insufficient for assembly, there exists of a largest value $k_0$ at which nearly all the $k$-mers in the reference genome are present in the reads for $k\leq k_0$. Thus, decreasing $k$ below $k_0$ only contributes to making more $k$-mers repeated (effect (ii)).

From these observations, we conclude that the number of distinct genomic $k$-mers in the reads is likely to reach a maximum value. At this value, all the $k$-mers in the reference genome are likely to appear in the reads, thus the assembly at this $k$-mer length nearly covers all the genome. Also, the assembly is likely to be of high contiguity as a large number of distinct $k$-mers imply that more repetitions are fully spanned by $k$-mers.

\end{revision}

\section{Results}

\subsection{Datasets and assemblers}
We benchmarked \ourtool using data from the Genome Assembly Gold-standard Evaluation (GAGE),
which was previously used to evaluate and compare different assemblers~\citep{GAGE}.
We used three datasets from three genomes of different sizes: \aureus (2.8 Mbp), human chromosome 14 (88 Mbp) and \bombus (250 Mbp). 
The datasets contain 5mil/62mil/497mil Illumina reads of length 101bp/101bp/124bp (respectively).
The coverages are 167x/70x/247x.

For each dataset, the GAGE study published the assembler that produced the best results,
along with its most effective formula (including commands and parameters).
To assess how our predicted $k$ values relate to the quality of assemblies, 
we choose, for each dataset, the de Bruijn assembler and formula
that produced the best results in the GAGE study.
For \aureus and \hchr, this was Velvet 1.2.08~\citep{Velvet} with the parameters given on the GAGE website. 
For \bombus, we used SOAPdenovo2~\citep{SOAPdenovo2} using the GAGE recipe and no additional parameters.
All experiments were run on a 32-core machine (Xeon E7-8837 @ 2.67 GHz) with 512GB RAM.

\subsection{Performance of \ourtool's choice of $k$ }
We ran \ourtool on all three datasets with putative $k$ values of $21,31,41,51,61,71,81$ and a sampling frequency of
$\epsilon = 1000$.
The optimal values of $k$ were predicted to be 31 for \aureus, 71 for \hchr, and 51 for \bombus.
We then assembled each dataset using both the optimal and other reasonable values of $k$.
The results of each assembly are shown in Table~\ref{tbl:assembly}.
The decision of what constitutes a ``best'' assembly is complicated due to the inherent 
trade-offs~\citep{GAGE, Assemblathon1, Assemblathon2}. 
We therefore evaluated each assembly using  three common quality measures: the contig NG50 length, the assembly size, and 
the number of assembly errors. 
Contig NG50 is defined as the length at which half of the predicted genome size is contained in contigs longer than this length. 
Contigs were obtained by splitting reported scaffolds at each undetermined nucleotide.
The size was measured as the sum total length of contigs larger than $500$ bp.
The Error column reflects the number of mis-joins called by the QUAST software~\citep{QUAST}.
Note that for \bombus there is no reference available and hence it is not possible to measure the number of errors.
QUAST reports an assembly error as a position in the assembled contigs where one of the following mis-assembly events occur: (i) the left flanking sequence aligns over 1 kbp away from the right flanking sequence on the reference, or (ii) they overlap by $>1$ kb, or (iii) the flanking sequences align on opposite strands or different chromosomes.


\begin{table}
\begin{tabular*}{.97\linewidth}{l l l l}
\toprule
Assembly & Contig NG50 (Kbp) & Size (Mbp) & Errors  \\
\midrule
\multicolumn{2}{l}{\aureus (Velvet)} & & \\
$k = 21$ & 0.5  & 7.65 & 0 \\
\underline{$k = 31$} & \textbf{19.4}  &  \textbf{2.83} & 10 \\
$k = 41$ & 11.7  & 2.81 &  \textbf{6} \\
$k = 51$ & 4.6  & 2.80 & 9 
\vspace{5pt} \\
\multicolumn{2}{l}{\hchr (Velvet)} & &\\
$k = 41$ & 2.4  & 74.56 & 764 \\
$k = 51$ & 4.0  & 79.92 & 843 \\
$k = 61$ & \textbf{5.4}  & \textbf{82.10} & 431 \\
\underline{$k = 71$} & 4.7  & 81.89 & 251 \\
$k = 81$ & 1.8  & 74.18 & \textbf{153} 
\vspace{5pt} \\
\multicolumn{2}{l}{\bombus (SOAPdenovo2)} & & \\
$k = 41$ & 5.4  & 224.05 &  \\
\underline{$k = 51$} & \textbf{10.4} & 229.71 &  \\
$k = 61$ & 9.5  & \textbf{230.36} &  \\
$k = 71$ & 5.9  & 226.11 &  \\
$k = 81$ & 2.5  & 207.11 &  \\
\midrule
\end{tabular*}
\caption{Quality of assemblies for different values of $k$. 
The value of $k$ predicted by \ourtool is underlined.
\label{tbl:assembly}}
\end{table}

For \aureus, the assembly with the chosen $k$ had the best NG50 and size, but had more errors than other assemblies. 
For \hchr, the assembly with the chosen $k$ did not have the best NG50 or size (though it was close), 
but did have significantly less errors.
For \bombus, the chosen $k$ gave an assembly with the best NG50 and second best size 
(the number of errors  is unknown since there is no reference).
Overall, \ourtool's choice of $k$ led to the best assemblies of \aureus and \bombus, as measured by 
the NG50 and assembly size, and to a good assembly of \hchr that represents a compromise between 
NG50/assembly size and the number of errors.

\revisedtext{\ourtool is multi-threaded, it uses one thread per $k$ value. For a single thread, the running time and memory usage of \ourtool is shown in Table~\ref{tbl:performance}.
We also compared the speed of our approximate histogram generation to what could be achieved by the exact \kmer counting method DSK~\citep{DSK}.
\ourtool ran 6-10 times faster than DSK, confirming that our sampling approach leads to significant speed-ups.
For the largest dataset (\bombus), the total wall-clock time of \ourtool with $k$-mer set $\{21, 31, 41, 51, 61, 71, 81\}$ using 7 threads is 3 hours.}
Note that assembling \bombus using SOAPdenovo2 requires approximately 30 CPU hours for a single value of $k$ (10 wall-clock hours, 8 threads).

\begin{table}
\begin{tabular*}{.97\linewidth}{l l l p{.15\textwidth}}
\toprule
Organism &  \multicolumn{2}{c}{CPU time} & Memory usage of  \\
\cline{2-3}
 & DSK & \ourtool & \ourtool (GB) \\
\midrule
\aureus & 2min & 11sec & 0.1 \\
\hchr & 48min & 7min & 0.1  \\ 
\bombus & 7.5hour & 1.2hour & 0.4 \\ 
\midrule
\end{tabular*}
\caption{
\revisedtext{
Resource utilization of \ourtool compared to a \kmer counting based approach (DSK).
We executed \ourtool and DSK for a single value of $k$ (81) using one thread. 
\ourtool was executed with a sampling frequency of $\epsilon = 1000$. DSK used $5$ GB of memory.
}  
\label{tbl:performance}
}
\end{table}

\subsection{Comparison with VelvetOptimizer and VelvetAdvisor}
We are aware of only two other methods that have been proposed to optimize $k$.
VelvetOptimizer (abbreviated VO) is an unpublished tool that attempts to
optimize $k$ by performing a Velvet assembly for each odd $k$ value between 19 and 79 and picking the one
that yields the highest scaffold N50
    (Seemann, \url{http://dna.med.monash.edu.au/~torsten/velvet_advisor}).
It then  determines coverage cut-off parameters that yield the longest assembly in contigs longer than 1 kbp. 
Since this requires doing 30 assemblies, using VO on the \hchr data would require close to half a CPU-year.
We therefore were not able to evaluate VO on \hchr or the even larger \bombus.


We did execute VO on the \aureus dataset, using the default optimization parameters.
Each Velvet assembly requires around 40 minutes of CPU time, resulting in 20 hours computation 
(though this can be parallelized).
VO selected the assembly with $k=41$, with expected coverage value of $12$, and cut-off value of $6.47$. 
Compared to the assembly based on \ourtool's choice of $k=31$ (shown in Table~\ref{tbl:assembly}), 
VO's assembly has slightly higher size (2.85kbp vs 2.83kbp), significantly lower contig NG50 (11.6kbp vs 19.4kbp), and significantly more errors (23 vs 10).
The VO assembly has a higher scaffold N50 (734 kbp versus 257 kbp), but this high N50 value may be misleading: QUAST reports a NGA50 value (corrected scaffold NG50) of 11.6 kbp for the VO assembly, and 110.2 kbp for \ourtool assembly. 
We conclude that \ourtool's choice of $k$ leads to a better assembly than VO's in this case.
We note, however, that we believe the main advantage of \ourtool over VO's approach 
is that it is orders of magnitude faster (2mins vs 20 hours) and is applicable even when VO is 
not feasible (e.g. \hchr and \bombus).


Another method is the unpublished tool Velvet Advisor
(Gladman and Seemann, \url{http://bioinformatics.net.au/software.velvetoptimiser.shtml}).
It uses a formula to recommend a value of $k$, given the number of reads, the read length and the estimated genome length. 
Velvet Advisor recommends $k=81$ for the \aureus dataset, $k=51$ for the \hchr dataset and $k=85$ for the \bombus dataset. 
The $k$ value for the \hchr dataset leads to a good assembly, 
but, for \aureus and \bombus, the assemblies using these values are very poor.

\subsection{Effect of sampling and the fit of the statistical model}
\begin{figure*}
    \begin{center}
	\subfigure[]{
	\includegraphics[width=.32\textwidth]{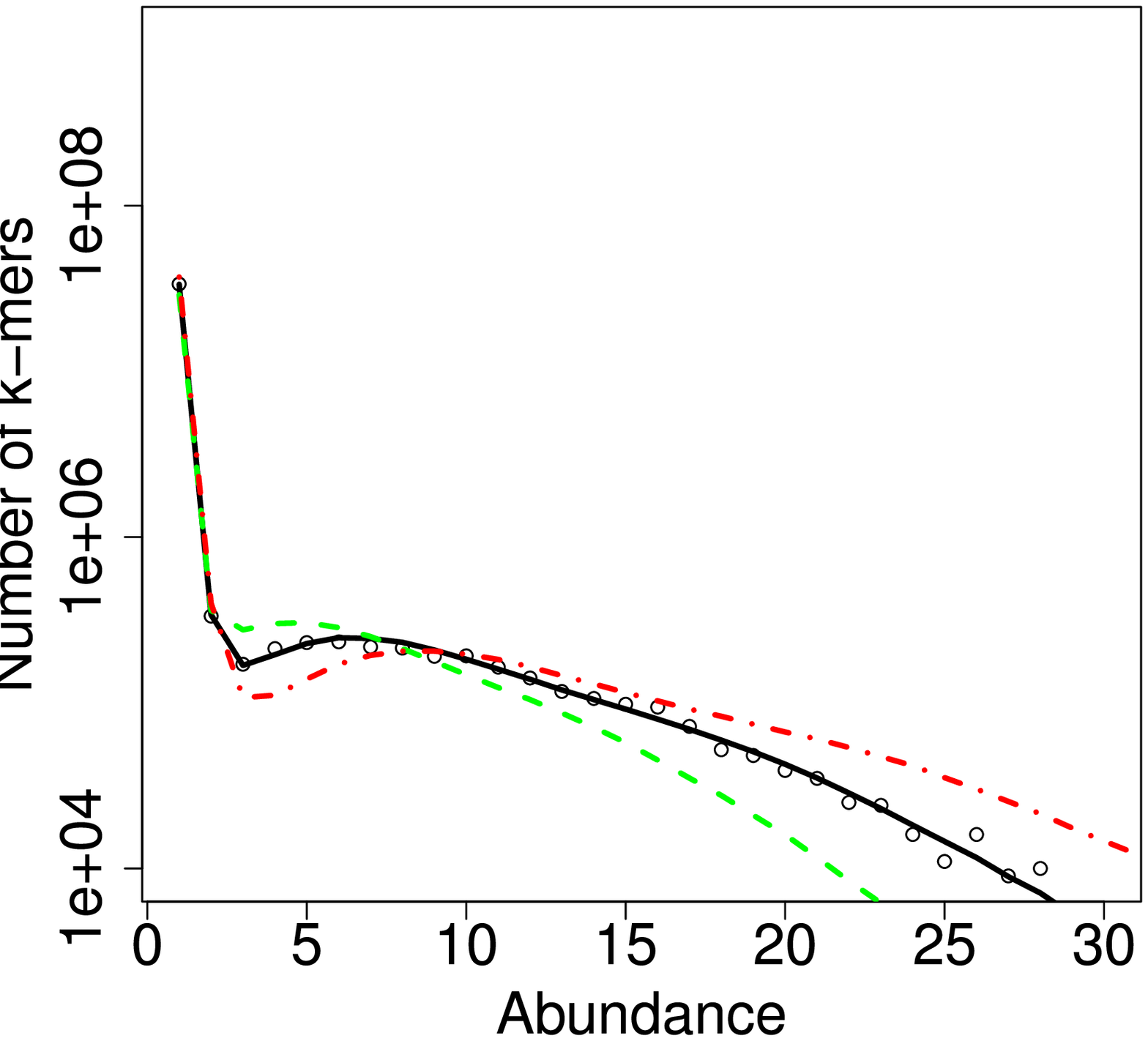}
    }
	\subfigure[]{
	\includegraphics[width=.32\textwidth]{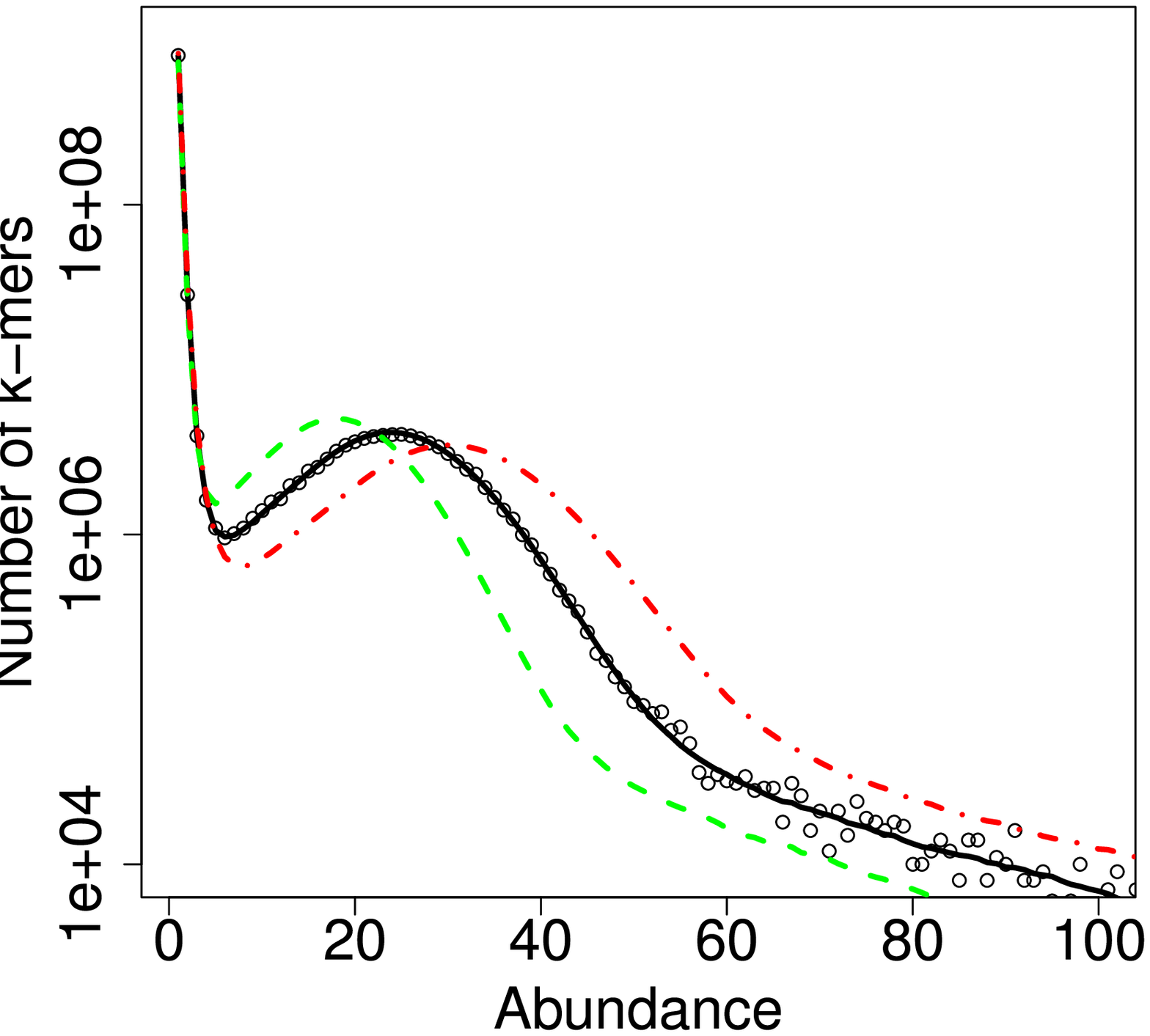}
	}
    \subfigure[]{
    \includegraphics[width=.32\textwidth]{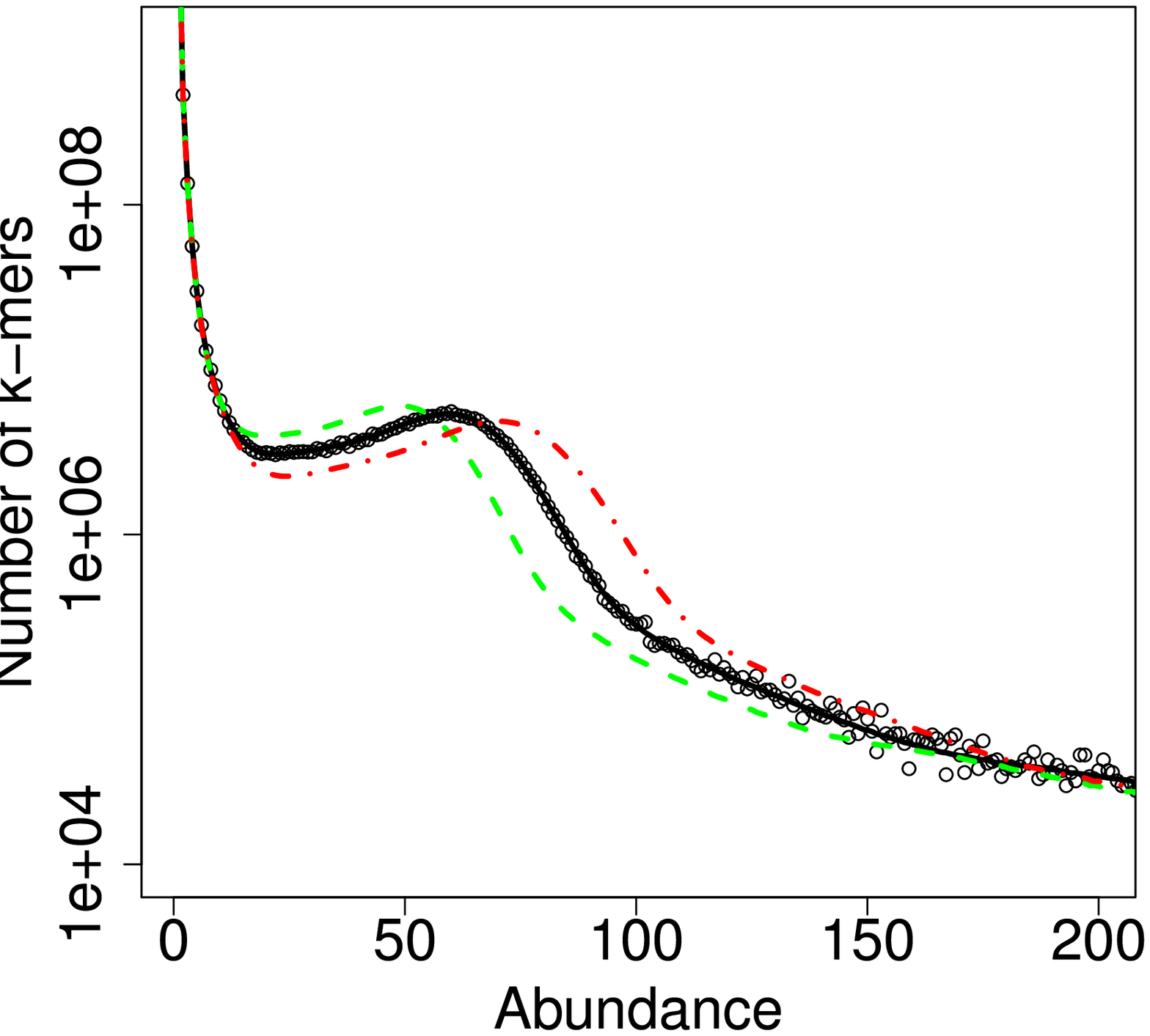}
	}
	\caption{\label{fig:sampling}
The accuracy of the sampling method. 
The panels reflect the three datasets: \aureus (a), \hchr (b), and \bombus (c).
Each plot show the exact histogram curves for $k=51$ (solid black curve), $k=41$ (dash-dot red curve),
and $k=61$ (dashed green curve).
The approximate (sampled) histogram is shown using black dots.
Note that $y$ is shown on a log-scale, exaggerating the differences at lower y values.
}
	\end{center}
\end{figure*}

Since the main purpose of the histograms is to contrast the differences between different $k$ values,
we measure the accuracy of our approximate histogram by comparing it at a fixed $k$ value (51)
to the exact distribution of $k=51$ and the exact distributions of nearby $k$ (41 and 61).
The results for our three datasets are shown in Figure~\ref{fig:sampling}.
We observe that the sampled histogram closely follows the exact one and
easily discriminates between other $k$ values when such a discrimination is possible from the exact counts.

\begin{figure*}
\includegraphics[width=\textwidth]{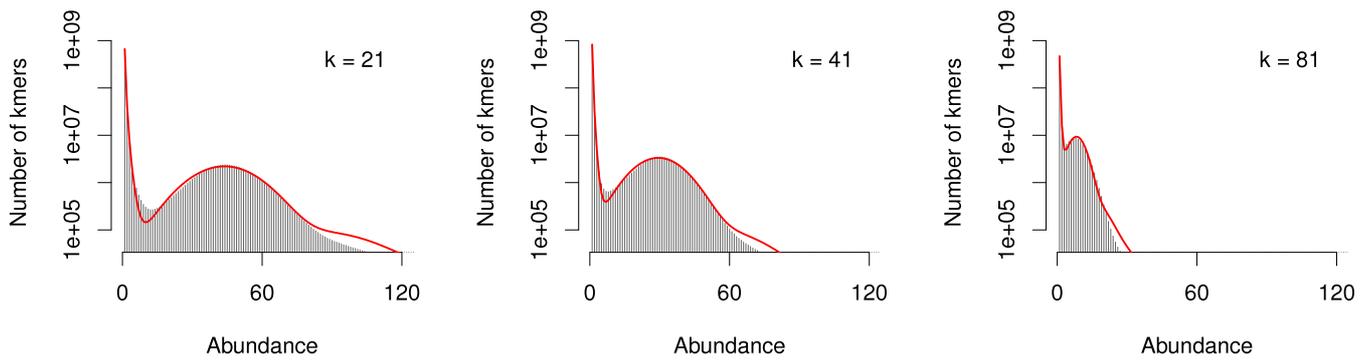}
\caption{
The abundance histograms for \hchr with $k$ values of 21, 41, and 81 (on a y log scale).
Each plot also shows a curve corresponding to the optimized statistical model (haploid).
\label{fig:humanhistogram}}
\end{figure*}

We illustrate the effect of $k$ on the abundance histogram for \hchr in Figure~\ref{fig:humanhistogram}.
In this case, the histogram is dominated by a mixture of a  distribution
for erroneous \kmers and one for genomic \kmers.
As $k$ increases, the genomic distribution shifts left and becomes more narrow,
resulting in a larger overlap  with the erroneous distribution.

\begin{figure*}
\includegraphics[width=0.66\textwidth]{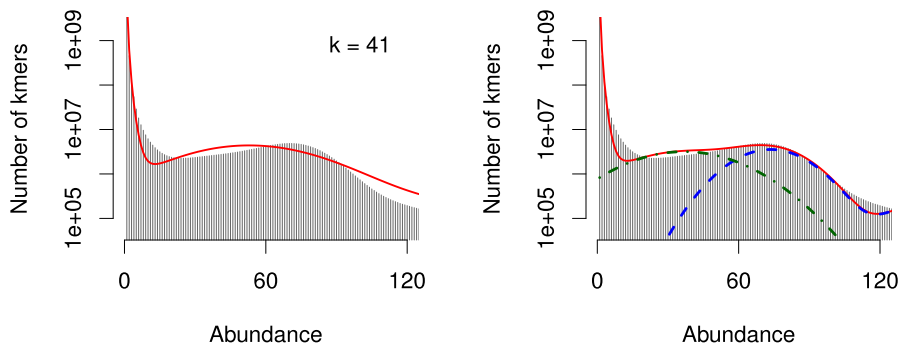}
\caption{
The abundance histogram and optimized model for \bombus, $k=41$, using a haploid  (left) and a diploid model (right), on a y log scale.
In both graphs, the black histogram curves are the actual \kmer histogram, and the red (solid) curve is the maximum likelihood fit using our model. 
In the diploid model graph, the green (dot-dashed) curve models the heterozygous \kmers, and the blue (dashed) curve models the homozygous \kmers.
Other components of the mixture are not shown.
\label{fig:bombushistogram}
}
\end{figure*}

Figure~\ref{fig:humanhistogram} also shows the fit of our model to the histogram.
Even though the human genome is diploid, its heterozygosity rate is small enough that we model the \kmer
abundance histogram using the haploid statistical model.  The high copycounts appear to be inaccurately fitted, but note that the log-scale amplifies the difference in low abundances.
For \bombus, on the other hand, the haploid model does not lead to a good fit, 
likely due to a possibly higher polymorphism rate.
Figure~\ref{fig:bombushistogram} shows the difference in fit between using a haploid and diploid model
for \bombus.

\subsection{Relation of the number of distinct genomic \kmers to assembly quality}

\begin{figure*}
\includegraphics[width=\textwidth]{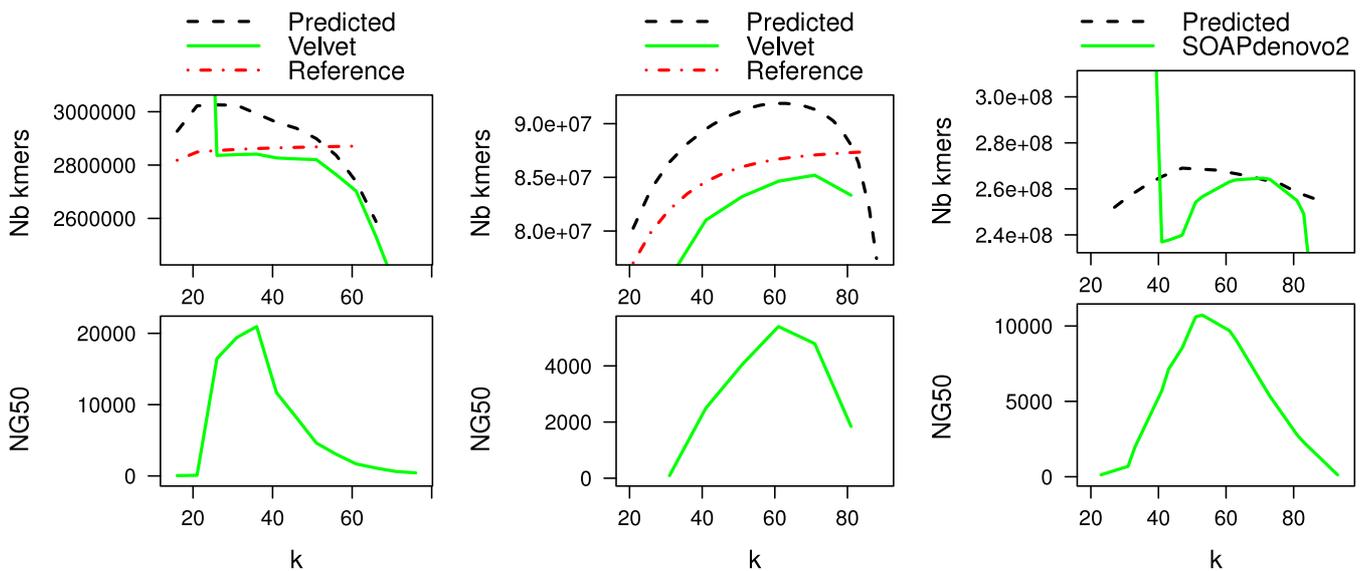}
\caption{
Relation of the number of distinct genomic \kmers to assembly quality.
We show the results for the three datasets: \aureus (left), \hchr (middle), \bombus (right).
We plot the number of distinct genomic \kmers predicted from the histogram from our model, 
the number present in the reference, and the number present in the assembly. 
We also show the NG50 of the assembly.
\label{fig:genomickmers}}
\end{figure*}

An important component of our method is the prediction of the number of distinct genomic \kmers 
from the abundance histogram.
Our underlying assumption is that providing the assembler with more distinct genomic \kmers 
leads to more of these \kmers being used in contigs and to longer contigs.
To measure this effect, we plot (as a function of $k$) the number of distinct genomic \kmers predicted from the histogram, 
the number of distinct \kmers used in the assembly, the NG50 of the assembly, and 
the number of distinct \kmers in the reference (Figure~\ref{fig:genomickmers}).

For \aureus and \hchr, the number of distinct genomic \kmers in the assembly approximatively mirrors the number of ones predicted in the input, with the exception of extreme $k$ values. 
For \bombus, the variations of the predicted number of $k$-mers do not match the variations of the number of distinct $k$-mers present in the assemblies. We postulate that this discrepancy may be due to heterozygosity, and note that the agreement between NG50 and our prediction is sufficient for our purpose.

For all three organisms, the NG50 rises and falls in accordance with the number of predicted distinct genomic \kmers.
We also observe that \ourtool overestimates the number of distinct genomic \kmers when compared to the reference $k$-mers.
A part of this is likely due to  heterozygosity, which is not captured in the haploid reference.
However, it may also partially indicate room for improvement in our statistical model and/or optimization.

For the lowest values of $k$ in the \aureus and \bombus datasets, the assemblers produce a much larger assembly than expected. We conjecture that this is due to a mis-estimation in the assemblers of what constitutes an erroneous $k$-mer (since with low $k$-mer sizes, erroneous $k$-mer have higher abundance than with high $k$-mer sizes). We verified this conjecture in the \aureus dataset, by manually assembling \aureus with Velvet using $k=21$ and a larger coverage cut-off value forced to 7 (experimentally found). The new Velvet assembly size is $2.8$ Mbp, which is much closer to the reference size than the $7.65$ Mbp assembly with automatic coverage cut-off. Thus, this indicates that larger assemblies are artifacts made by assemblers rather than an actual increase of genomic $k$-mers.

\section{Discussion}

While we have presented a method that attempts to find the best value of $k$ for assembly, 
we would like to note several limitations inherent in this approach.
First of all, \ourtool may in some instances report that a best value of $k$ cannot be found 
because it is not able to fit the generative model to the abundance histograms.
This could simply be due to a limitation of our model or the optimization algorithm, 
but it could also be due to a difficulty inherent in the data.
For example, data from single cell experiments have uneven coverage~\citep{Chitsaz11},
violating a basic assumption of our model.
Similarly, data from metagenomic or RNA-seq experiments do not come from a single genome and
their histograms have different properties. 
In these cases, it has been observed that there is often no single best $k$ and
and that combining the assemblies from different $k$ can be beneficial. 
Though  \ourtool does not suggest a $k$ in these cases, 
it  provides the abundance histograms that can be useful in determining the best assembly approach.

We have demonstrated our approach to be useful for de Bruijn based assemblers.
Other assemblers, such as SGA~\citep{SGA}, follow the alternate string overlap graph approach,
in which reads are not chopped up into \kmers.
These assemblers do not have the $k$ parameter but do have an alternate parameter for the minimum length of a non-spurious overlap.
Though a formal relation between these parameters has not been established, they play a  similar role
in affecting the assembly results. 
We therefore consider it an interesting direction for future research to extend our approach 
to select the best overlap parameter for string overlap graph assemblers.

Finally, we wish to emphasize that our benchmark did not attempt to produce the best possible assembly for each organism. 
Rather, we are restricting ourselves to what a ``typical'' user might do with the data: 
run a single assembly software on un-corrected data, and possibly try several $k$-mer values. 
In order to get the best possible assembly, one would have to explore the Cartesian product of several assemblers, several read error-correction methods, and several $k$-mer values, 
which is often a prohibitively long task. 
Notably, because we did not select the best error-correction method for each assembler/organism, 
the assemblies reported in Table~\ref{tbl:assembly} have lower contig N50 
(and also scaffold N50, data not shown) 
than those reported in the GAGE benchmark.

\begin{revision}
There are improvements that we have left for future work.
The first direction is to determine how our method could be applied to non-uniform coverage. While a single best $k$ value for metagenome and transcriptome assembly is unlikely to exist, perhaps useful information could be extracted from the histograms constructed on such datasets.
The second direction is to explore ways of improving the accuracy of our statistical model, potentially leading to more accurate estimates of the number of distinct genomic \kmers.
\end{revision}

\section*{Acknowledgements}
We would like to thank Qunhua Li and Francesca Chiaromonte for useful discussions.

\bibliographystyle{plainnat}
\bibliography{bibliography}

\end{document}